\documentclass[conference]{IEEEtran}

\pdfoutput=1

\ifCLASSINFOpdf
\else
\fi
\usepackage[cmex10]{amsmath}

\usepackage{graphicx,epic,eepic,epsfig,amsmath,latexsym,amssymb,mathrsfs,verbatim,subfigure,color}

\newtheorem{definition}{Definition}

\newtheorem{lemma}[definition]{Lemma}

\newtheorem{theorem}[definition]{Theorem}

\newtheorem{conjecture}[definition]{Conjecture}

\newcommand{\ket}[1]{|#1\rangle}

\def\squareforqed{\hbox{\rlap{$\sqcap$}$\sqcup$}}
\def\qed{\ifmmode\squareforqed\else{\unskip\nobreak\hfil
\penalty50\hskip1em\null\nobreak\hfil\squareforqed
\parfillskip=0pt\finalhyphendemerits=0\endgraf}\fi}
\def\endenv{\ifmmode\;\else{\unskip\nobreak\hfil
\penalty50\hskip1em\null\nobreak\hfil\;
\parfillskip=0pt\finalhyphendemerits=0\endgraf}\fi}

\long\def\ignore#1{}

\def\CC{\mathcal{C}}
\def\CR{\mathcal{Q}}
\def\CT{\mathcal{R}}
\def\CCOne{\mathcal{C}_{1}}
\def\CROne{\mathcal{Q}_{1}}
\def\CTOne{\mathcal{R}_{1}}
\def\Net{\mathcal{N}}
\def\Tensor{\boxtimes}
\def\Mul{\odot}
\def\QMC{\operatorname{MC}}
\def\T{\mathcal{T}}
\def\C{\mathbb{C}}

\begin{document}
%
\title{Quantum Capacities for Entanglement Networks}

\author{\IEEEauthorblockN{Shawn X Cui\IEEEauthorrefmark{1}, Zhengfeng Ji\IEEEauthorrefmark{2}\IEEEauthorrefmark{3}, Nengkun Yu\IEEEauthorrefmark{3}\IEEEauthorrefmark{4}, and Bei Zeng\IEEEauthorrefmark{3}\IEEEauthorrefmark{4}\IEEEauthorrefmark{5}}
\medskip
\IEEEauthorblockA{
\IEEEauthorrefmark{1}Department of Mathematics, University of
California, Santa Barbara, CA 93106, USA\\
\IEEEauthorrefmark{2}Centre for Quantum Computation \& Intelligent
Systems, School of Software,\\
Faculty of Engineering and Information Technology, University of
Technology, Sydney, NSW 2007, Australia\\
\IEEEauthorrefmark{3}Institute for Quantum Computing, University of Waterloo,
Waterloo,  Ontario, N2L 3G1, Canada\\
\IEEEauthorrefmark{4}Department of Mathematics $\&$ Statistics,
University of Guelph, Guelph, ON, N1G 2W1, Canada\\
\IEEEauthorrefmark{5} Canadian Institute for Advanced Research, Toronto,
  Ontario, M5G 1Z8, Canada}}


%


\maketitle


\maketitle

\begin{abstract}
  We discuss quantum capacities for two types of entanglement
  networks: $\CR$ for the quantum repeater network with free classical
  communication, and $\CT$ for the tensor network as the rank of the
  linear operation represented by the tensor network. We find that
  $\CR$ always equals $\CT$ in the regularized case for the same
  network graph. However, the relationships between the corresponding
  one-shot capacities $\CROne$ and $\CTOne$ are more complicated, and
  the min-cut upper bound is in general not achievable. We show that
  the tensor network can be viewed as a stochastic protocol with the
  quantum repeater network, such that $\CTOne$ is a natural upper
  bound of $\CROne$. We analyze the possible gap between $\CTOne$ and
  $\CROne$ for certain networks, and compare them with the one-shot
  classical capacity of the corresponding classical network.
\end{abstract}



\section{Introduction}
\label{sec:intro}

The study of quantum information transmission via a quantum repeater
network is of both theoretical and practical
relevance~\cite{briegel1998quantum,kimble2008quantum,satoh2012quantum,duan2010colloquium}.
Given an undirected graph $G = (V,E)$ with a dimension function
$d: E \rightarrow \mathbb{N}_{\geq 2}$ and a set of sources
$S \subset V$ (resp. sinks $T \subset V$), a quantum repeater network
associated to $(G,d,S,T)$ can be viewed as an entanglement network,
where each pair of connected nodes of $e \in E$ share a maximally
entangled state
$\ket{\psi_{e}}=\sum_{i=1}^{d_{e}}\ket{ii}/\sqrt{d_{e}}$. Quantum
information is transmitted from some source vertex $s_i \in S$ to some
sink vertex $t_j \in T$ through the network via local quantum
operations and classical communications. In this work we focus on
networks with one source vertex $s$ and one sink vertex $t$. The
general case for transmitting information
from $S$ to $T$ reduces to this simple case by viewing all $s_i$/$t_j$
as one source/sink.

Usually, the dimension $d_e$ of the maximally entangled state $\ket{\psi_{e}}$ is
chosen to be the same on each edge $e$. However, the more general case, where the $d_e\,'$s may be different, is also known to be interesting, which has demonstrated connections to the topological quantum field theory and the theory of quantum gravity~\cite{cui2015quantum,ryu2006holographic,headrick2014causality,HNQ+16}. In this work, we will discuss the general case where $d_e\,'$s may be different, and show that this `inharmony' of these dimensions may have interesting effects on the one-shot capacities of the corresponding networks.

\begin{figure}[htbp]
\centering
  \includegraphics[width=3.0in]{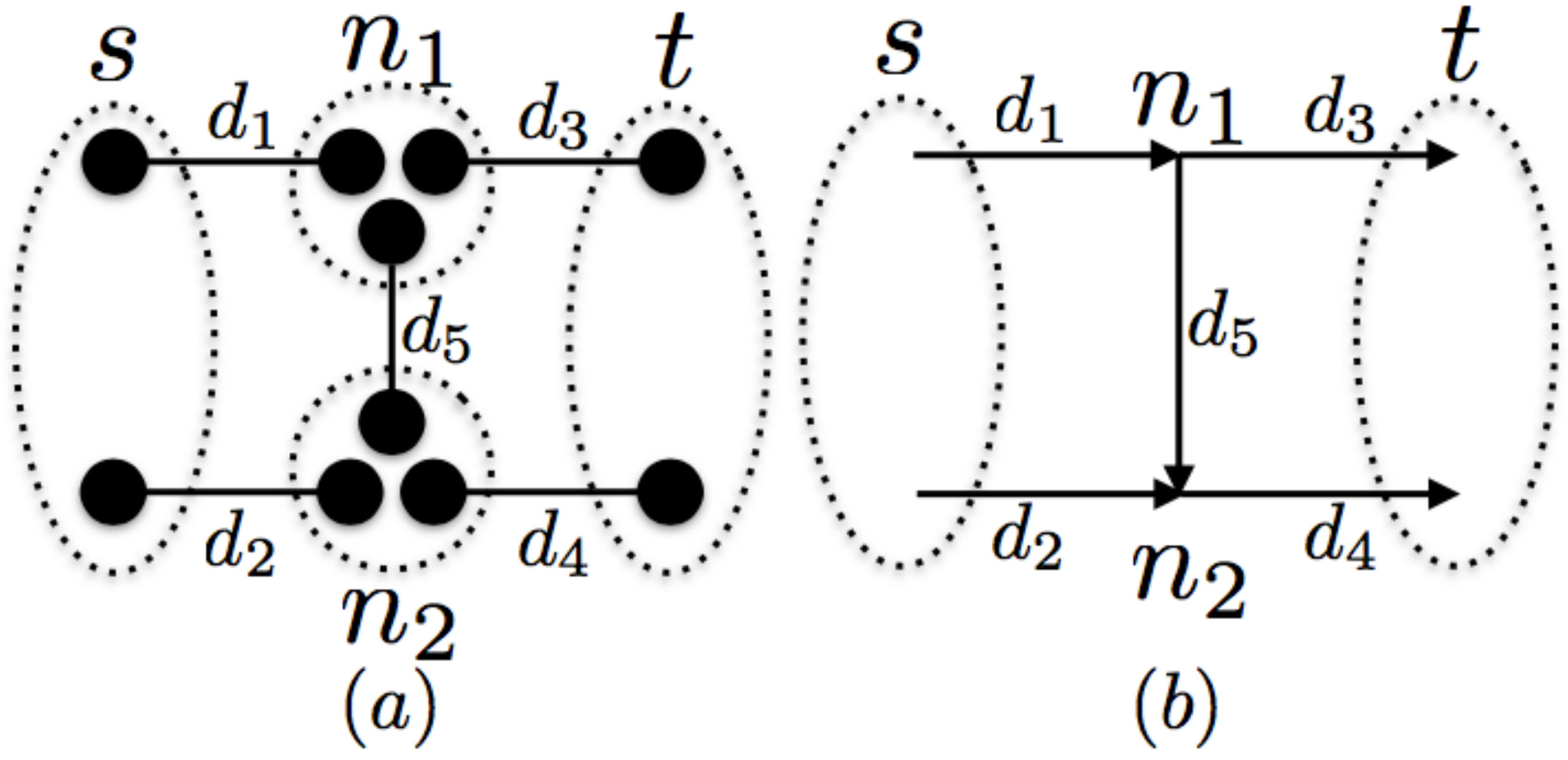}
  \caption{(a) A repeater network with four vertices (a source $s$, a sink $t$
  and two nodes $n_1$, $n_2$) and five edges (each associated with
  a maximal entangled state of dimension $d_i$).  (b) A corresponding quantum network with a directed graph. The directed edges represent noiseless quantum channels.}
  \label{fig1}
\end{figure}

An example of quantum repeater network is shown in Fig~\ref{fig1}(a). The graph $G$ has four vertices (a source $s$, a sink $t$ and two nodes $n_1$, $n_2$), and five edges with dimensions $d_i$ ($i=1,2,3,4,5$) respectively.

We consider the case where arbitrary quantum operations are allowed at each vertex, and classical communications are free (other kinds of quantum networks are also considered in literature, see e.g. ~\cite{nishimura2014quantum} and references therein). In other words, we allow protocols given by local operations and classical communications (LOCC). The goal is then to establish maximum bipartite entanglement between $S$ and $T$ via LOCC. The capacities of this network can hence be defined accordingly.

\begin{definition}
Given the network $\mathcal{N} = (G, d, S, T)$, the one-shot capacity $\CROne(\Net)$ of the quantum repeater network associated with $\Net$ is given by the maximum dimension $d_{\max}$ of the maximally entangled state that can be created between $S$ and $T$ via LOCC. The capacity $\CR(\Net)$ of this network is the regularized version of $\CROne(\Net)$,
\begin{equation}
\CR(\Net)=\lim_{n\rightarrow\infty}\left[\CROne(\Net^{\Tensor n})\right]^{1/n},
\end{equation}
where $\Net^{\Tensor n}$ is the network $(G,d^n,S,T)$.
\end{definition}

We remark that usually we should view `$\log_2\CR(\Net)$' as the capacity in terms of `bits'. For the discussion of
this work, we simply omit the `$\log_2$' and directly use the dimension $\CR(\Net)$ as the capacity.

Since we allow unlimited classical communications, the capacity $\CR(\Net)$ is in fact the same as any quantum network, where each maximally entangled state associated with an edge $e$ is replaced by a directed quantum channel with an arbitrarily chosen direction and this channel has the capacity to transmit a quantum state of dimension $d_e$~\cite{hayashi2007quantum,leung2010quantum,jain2011quantum,satoh2012quantum}. Denote by $\tilde{G}$ a directed graph obtained from the quantum network corresponding to the quantum repeater $(G,d,S,T)$. So $\tilde{G}$ is the same as $G$ when the directions are ignored. Notice that the direction of each edge does not matter for the values of $\CR$/$\CROne$, since one can always reverse the direction by quantum teleportation. In other words, for any network $\tilde{\mathcal{N}}=(\tilde{G}, d, S, T)$, the quantum capacities are the same as the network $\mathcal{N}=({G}, d, S, T)$. For example, one of the corresponding quantum networks of the quantum repeater network in Fig~\ref{fig1}(a) is given in Fig~\ref{fig1}(b), where the graph is directed arbitrarily.


For a directed graph $\tilde{G}$ that corresponds to $G$, we also define the classical capacities $\CC$/$\CCOne$ of the network $\mathcal{N} = (\tilde{G}, d, S, T)$ as below. Different from the quantum case, the choice of directions of the edges may effect values of $\CC$/$\CCOne$, in general.

\begin{definition}
Given the network $\tilde{\Net}= (\tilde{G}, d, S, T)$, the one-shot capacity $\CCOne(\tilde{\Net})$ of the classical network associated with $\tilde{\Net}$ is given by the maximum cardinality $l_{\max}$ of the alphabet that can be transmitted from $S$ to $T$ via network coding. The capacity $\CC(\tilde{\Net})$ of this network is hence the regularized version of $\CCOne(\tilde{\Net})$, i.e.
\begin{equation}
\CC(\tilde{\Net})=\lim_{n\rightarrow\infty}\left[\CCOne(\tilde{\Net}^{\Tensor n})\right]^{1/n},
\end{equation}
where $\tilde{\Net}^{\Tensor n}$ is the network $(\tilde{G},d^n,S,T)$.
\end{definition}

Another interesting type of entanglement networks we focus on is the tensor network, which transports linear algebraic things like rank and entanglement~\cite{bauer2011tensor,cui2015quantum}. An example of the tensor network is given in Fig~\ref{fig2}. Given a network $\Net = (G, d, S,T)$, let $\tilde{V} = V \setminus (S \sqcup T)$, and for each $v \in V$, let $E(v) \subset E$ be the set of edges containing $v$. Also define $V(S) = \bigotimes\limits_{e \in E(v), v \in S} \C^{d_e}$, and define $V(T)$ analogously. One can assign a set of tensors $\T = \{\T_u: u \in \tilde{V}\}$ to $\Net$, where $\T_u$ is an arbitrary tensor in $\bigotimes\limits_{e \in E(u)}\C^{d_e}$, i.e., each index of $\T_u$ corresponds to an edge containing $u$. Then contracting the tensors along all internal edges results in a linear map $\beta_{\T}: V(S) \rightarrow V(T)$. The maximal rank of $\beta_T$ is considered to be the capacity of the tensor network. Explicitly, it is defined as follows:
\begin{figure}[htbp]
\centering
  \includegraphics[width=1.5in]{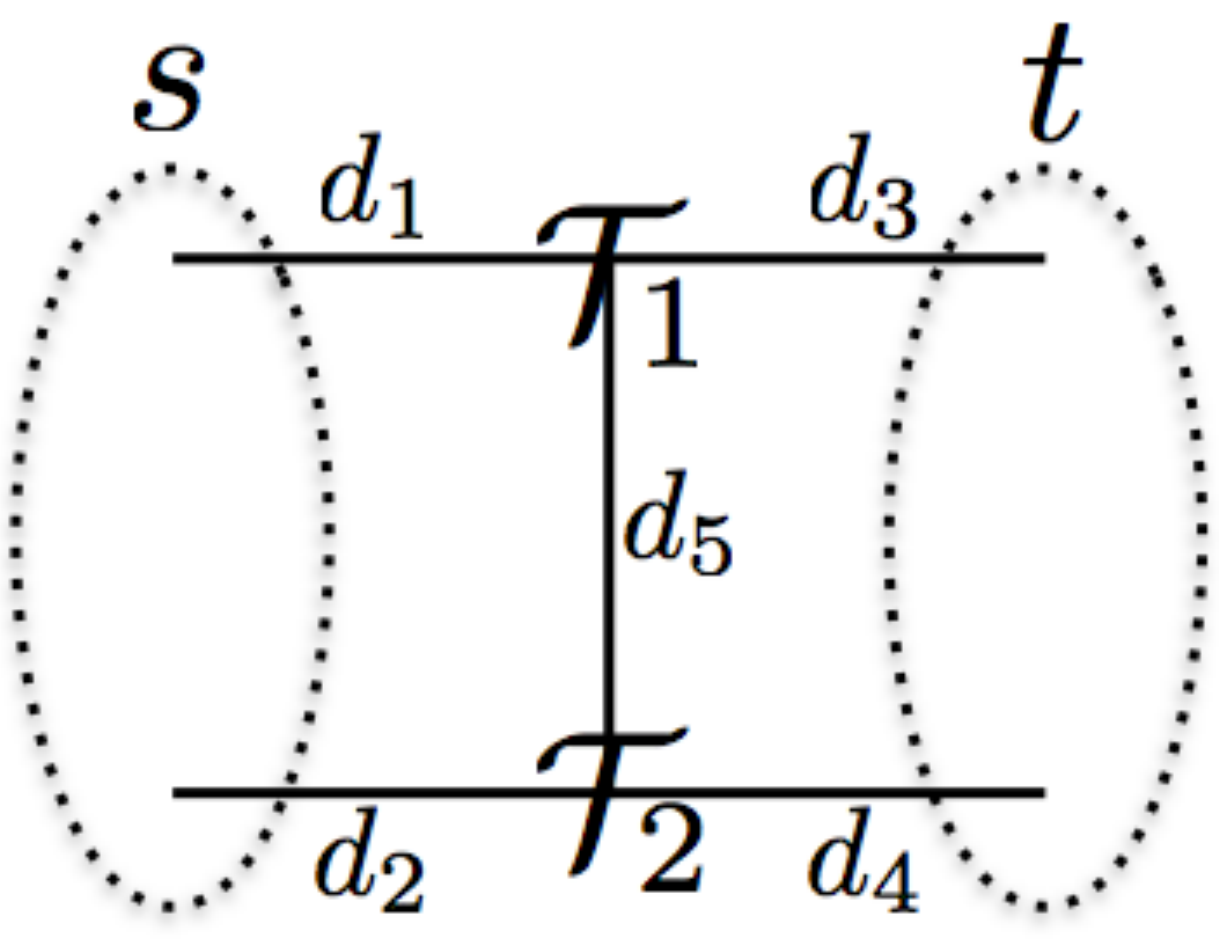}
  \caption{A tensor network}
  \label{fig2}
\end{figure}

\begin{definition}
The one-shot capacity $\CTOne(\Net)$ associated with $\Net$ is defined to be the maximal rank of $\beta_{\T}$ over all tensor assignments $\T$. And similarly, the capacity $\CT(\Net)$ is defined as the regularized version of $\CTOne(\Net)$, namely,
\begin{equation}
\CT(\Net)=\lim_{n\rightarrow\infty}\left[ \CTOne(\Net^{\Tensor n}) \right]^{1/n}.
\end{equation}
\end{definition}

This work studies $\CR$/$\CROne$, $\CC$/$\CCOne$, $\CT$/$\CTOne$, and their relationships.

\section{The min-cut upper bound}
\label{sec:reg}

We start from the natural upper bound for all the capacities given by the min-cut of the graph.

\begin{definition}
  For a network $\Net = (G, d, S,T)$ with $G = (V,E)$ an undirected graph, a cut $C = (\tilde{S}, \tilde{T})$ is a partition $V = \tilde{S} \sqcup \tilde{T}$, such that $S \subset \tilde{S}$ and $T \subset \tilde{T}$. The min-cut, $\QMC(\Net)$, associated with $\Net$ is defined as the minimum, over all possible cuts $C = (\tilde{S}, \tilde{T})$, of the value $\prod\limits_{(u,v)\in E, u \in \tilde{S}, v \in \tilde{T}} d_{(u,v)}$.

 Similarly, for a network $\tilde{\Net} = (\tilde{G}, d, S,T)$ with $\tilde{G} = (V,E)$ a directed graph, $\QMC(\tilde{\Net})$ is defined in the same way as the undirected case except in the product above one only considers directed edges.
\end{definition}

It is obvious that for any directed graph $\tilde{G}$ corresponding to the undirected $G$, $\QMC(\tilde{\Net})\leq\QMC(\Net)$. And there always exists some $\tilde{G}$, which may
have directed cycles, such that $\QMC(\tilde{\Net})=\QMC(\Net)$. One such simple possibility is that we choose every edge of $\tilde{G}$ to be `bidirectional' (which
essentially corresponds to a directed cycle of length $2$).

Given the network $\tilde{\Net} = (\tilde{G}, d, S, T)$, it is well known that $\CCOne(\tilde{\Net})$ is upper bounded by the min-cut $\QMC(\tilde{\Net})$ of the graph. That is, $\CCOne(\tilde{\Net})\leq\QMC(\tilde{\Net})$. And this upper bound is achievable for $\CC$, i.e. $\CC(\tilde{\Net})=\QMC(\tilde{\Net})$, which is given by the famous max-flow/min-cut theorem~\cite{ahlswede2000network,Eugene2001,Papadimitriou1998,Vazirani2004}.

For the quantum network, one also naturally has $\CROne(\Net)\leq\QMC(\Net)$,  for both directed and undirected graphs. It is also known that for any directed acyclic graph, there is a quantum network coding protocol that simulates the classical network coding protocol on the same network~\cite{de2014quantum,kobayashi2009general,kobayashi2010perfect,kobayashi2011constructing}. This then gives $\CR(\Net)\geq\QMC(\tilde{\Net})$ for all directed acyclic graph $\tilde{G}$ such that $\tilde{\Net}=(\tilde{G},d,S,T)$.

Similarly, it is straightforward to show that $\CTOne$ is also upper bounded by the min-cut of the graph, i.e., $\CTOne(\Net)\leq \QMC(\Net)$. It is known that under certain circumstances, this upper bound is not achievable by $\CTOne$~\cite{cui2015quantum}. That is, there exists some network $\Net_0$, such that $\CTOne(\Net_0)$ is strictly smaller than $\QMC(\Net_0)$. It remains open whether $\QMC(\Net)$ is achievable by $\CT(\Net)$. We show $\QMC(\Net)$ is indeed also achievable by tensor networks, as given by the following theorem.

\begin{theorem}
\label{th:reg}
\begin{equation}
\CT(\Net)=\QMC(\Net)
\end{equation}
\end{theorem}

\begin{IEEEproof}
To prove this theorem, we begin with the following observations:

1. For two networks $\Net_l= (G, d^{(l)}, S,T)$, $\Net_u= (G, d^{(u)}, S,T)$ with the same underline graph but different dimension functions satisfying $d^{(l)}\leq d^{(u)}$ for all edges, we have $\CTOne(\Net_l)\leq \CTOne(\Net_u)$, and $\QMC(\Net_l)\leq \QMC(\Net_u)$.

2. $\QMC(\Net)$ is multiplicative, that is $\QMC(\Net)=\QMC(\Net^{\Tensor n})^{1/n}$.

3. $\CTOne(\Net)=\QMC(\Net)$ if the dimension on each edge is a power of some fixed integer $r$. This is Theorem $5.2$ of \cite{cui2015quantum}.

For any integer $n > 0$ and a quantum network $\Net= (G, d, S,T)$, we define two other networks $\Net_l= (G, d^{(l)}, S,T)$, $\Net_u= (G, d^{(u)}, S,T)$ corresponding to the network $\Net^{\Tensor n}$, with $d^{(l)}_e=2^{\lfloor n\log_2 d_e\rfloor}$, $d^{(u)}_e=2^{\lceil n\log_2 d_e\rceil}$ for any edge $e$. According to the first and the third observation, we have
\begin{eqnarray}
\QMC(\Net_l)&=&\CTOne(\Net_l)\leq \CTOne(\Net^{\Tensor n})\leq \CTOne(\Net_u)\nonumber\\
&=&\QMC(\Net_u).
\end{eqnarray}
On the other hand, notice that $d^{(l)}_e\geq d_e^n/2$ and $d^{(u)}_e\leq 2d_e^n$ for any edge $e$. Then,
\begin{eqnarray}
2^{-c_1}\QMC(\Net^{\Tensor n})&\leq& \QMC(\Net_l)\leq \CTOne(\Net^{\Tensor n})\nonumber\\
&\leq& \QMC(\Net_u)\leq 2^{c_2}\QMC(\Net^{\Tensor n}),
\end{eqnarray}
where $c_1$ and $c_2$ are the number of edges in the min cut of $\Net_l$ and $\Net^{\Tensor n}$, respectively, which are both bounded above by the total number of edges of $\Net$.

Then we can conclude that,
\begin{eqnarray}
2^{-c_1}\QMC(\Net^{\Tensor n})\leq \CTOne(\Net^{\Tensor n}) \leq 2^{c_2}\QMC(\Net^{\Tensor n}).
\end{eqnarray}
That is,
\begin{eqnarray}
\lim_{n\rightarrow\infty} 2^{-c_1/n}\QMC(\Net)&\leq& \lim_{n\rightarrow\infty}\left[\CTOne(\Net^{\Tensor n})\right]^{1/n}\nonumber\\
&\leq& \lim_{n\rightarrow\infty} 2^{c_2/n}\QMC(\Net).
\end{eqnarray}
Therefore,
\begin{equation}
\CT(\Net)=\QMC(\Net)
\end{equation}
\end{IEEEproof}

The following observation clarifies the relation between the quantum
repeater network and the tensor network.

\begin{lemma}
\label{lm:SLOCC}
For any network $\Net$, the tensor network is a stochastic local operation assisted by classical communication (SLOCC) protocol with the quantum repeater network.
\end{lemma}

\begin{IEEEproof}
Notice that for any network $\Net=(G,d,S,T)$, the tensor network can be obtained from a rank one projection $\Pi_i$
at each internal node $n_i$ (vertex of $G$ that is not a source or a sink), which is given by the so-called `projective entanglement pairs' (PEPS) representation of tensor networks~\cite{verstraete2008matrix}. Each rank one projection
$\Pi_i$ can be viewed as a local measurement with two outcomes $\{\Pi_i,I-\Pi_i\}$, where $I$ is the identity operator, with finite probability to obtain the measurement outcome $\Pi_i$.  This then corresponds to
an SLOCC protocol.
\end{IEEEproof}

\begin{theorem}
\label{corQ1}
For any network $\Net$, we have
\begin{equation}
\CROne(\Net)\leq\CTOne(\Net).
\end{equation}
And hence $\CR(\Net) \leq \CT(\Net)$.
\end{theorem}

\begin{IEEEproof}
We need to show some kind of converse of Lemma~\ref{lm:SLOCC}. If the one-shot
capacity of the network is $\CROne(\Net)=r$, then there is an LOCC protocol that
can distill a maximally entangled state of rank $r$ between the source and sink.
The stochastic version of such an LOCC protocol can be given by a linear transformation
on each node~\cite{dur2000three}. Since the output is a maximally entangled state, the SLOCC
can then be realized by a rank one projection on each node, which corresponds to a tensor network
as given by Lemma~\ref{lm:SLOCC}.
\end{IEEEproof}

We remark that a combination of the above theorem and results from~\cite{SVW05} gives rise to an alternative proof of the statement in Theorem~\ref{th:reg}.

Another viewpoint on Theorem~\ref{th:reg} is to
think about network coding as a protocol of measurement-based
quantum computation~\cite{de2014quantum}, which can then be realized
as a tensor network~\cite{gross2007measurement}.

\section{One-shot Capacities}
\label{sec:onshot}

The inequality $\CTOne(\Net)\leq\QMC(\Net)$ is known to be strict for some tensor networks.
For instance, see the tensor network in Fig.~\ref{fig2} with $d_1=d_4=5, d_2=d_3=3, d_5=2$. In this network, $\QMC(\Net)=15$, but $\CTOne(\Net)=14$,~\cite{cui2015quantum}.

Thus the min-cut upper bound $\QMC(\Net)$ is in general not achievable by the one-shot capacity $\CTOne(\Net)$ of tensor
networks. Theorem ~\ref{corQ1} then indicates that $\QMC(\Net)$ is also in general not achievable by the one-shot capacity $\CROne(\Net)$ of quantum repeater networks.

It remains open if it is always true that $\CROne(\Net)=\CTOne(\Net)$. We discuss the possible gap between them in this section. Before doing so, we first discuss how to simulate the classical network coding protocol for directed graphs with directed cycles.

\subsection{Directed graph with directed cycles}
\label{sec:cycle}


If the directed graph $\tilde{G}$ has cycles of length $2$, with some $d_i$ being a composite number, there is a way to change the
quantum repeater network into an equivalent network with directed acyclic graph by separating the vertices with cycles. To demonstrate our idea, we use the network given in Fig~\ref{fig1}(a) as an example, which can be
naturally generalized to other cases for directed graphs with directed cycles of length $2$.

\begin{figure}[htbp]
\centering
  \includegraphics[width=2.0in]{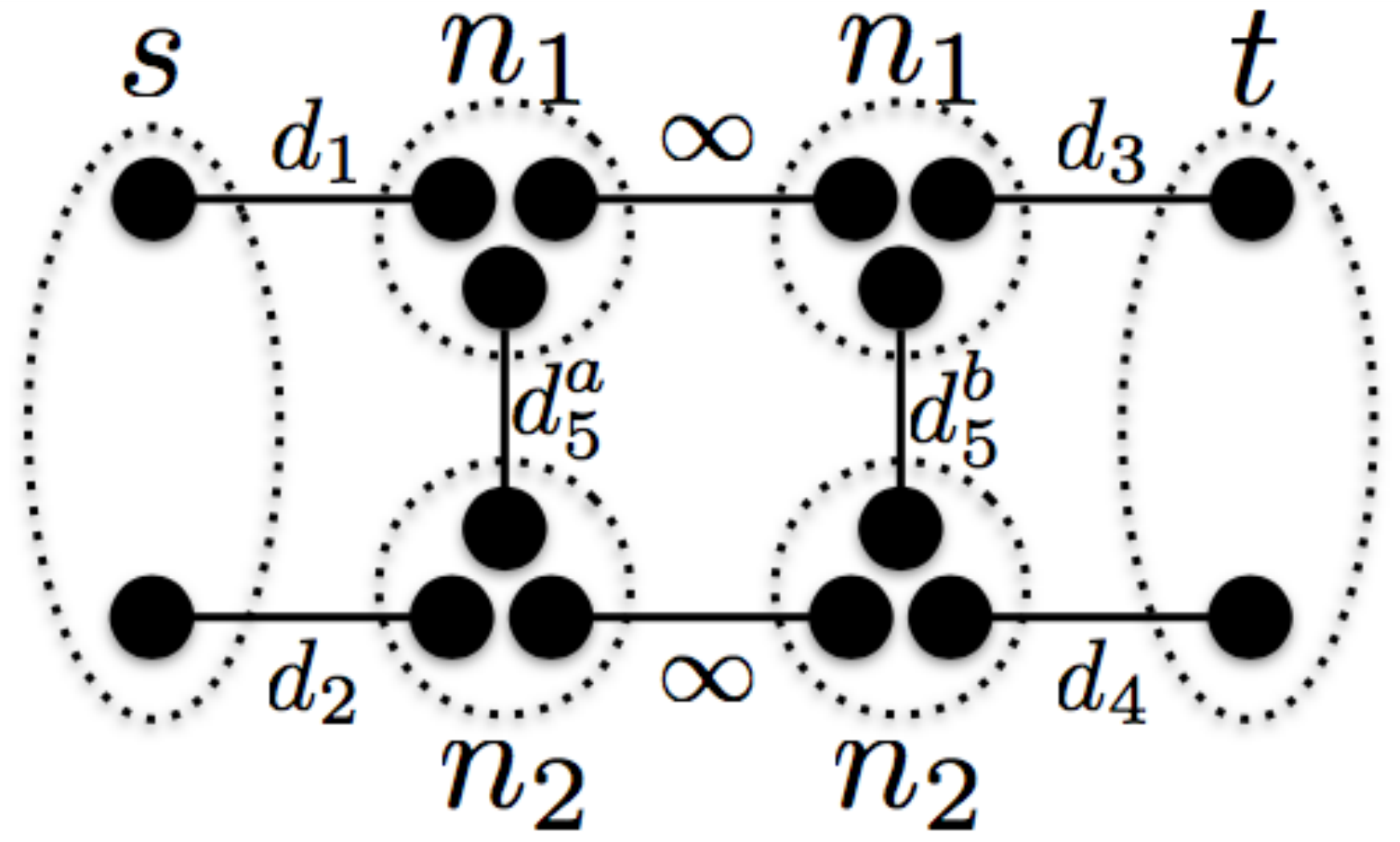}
  \caption{An equivalent repeater network of Fig.~\ref{fig1}(a) with $d_5^ad_5^b=d_5$. Each node ($n_1$ or $n_2$) is split into two nodes that are connected by a maximally entangled state with infinite dimension. The left two nodes of $n_1$ and $n_2$ are connected by a maximally entangled state with dimension $d_5^a$. The right two nodes of $n_1$ and $n_2$ are connected by a maximally entangled state with dimension $d_5^b$.}
  \label{fig3}
\end{figure}

As demonstrated in Fig~\ref{fig3}, where $d_5=d_5^ad_5^b$ for some $d_5^a>1, d_5^b>1$, we can split each node (e.g. $n_1$) into two nodes that are connected by a maximally entangled state with infinite dimension (i.e., the two nodes can transfer information from one to the other as they are essentially one node). Since $d_5=d_5^ad_5^b$, we can then think that one connection of $n_1$ and $n_2$ is a maximally entangled state with dimension $d_5^a$ and the other of $n_1$ and $n_2$ is a maximally entangled state with dimension $d_5^b$.

\begin{figure}[htbp]
\centering
  \includegraphics[width=3.5in]{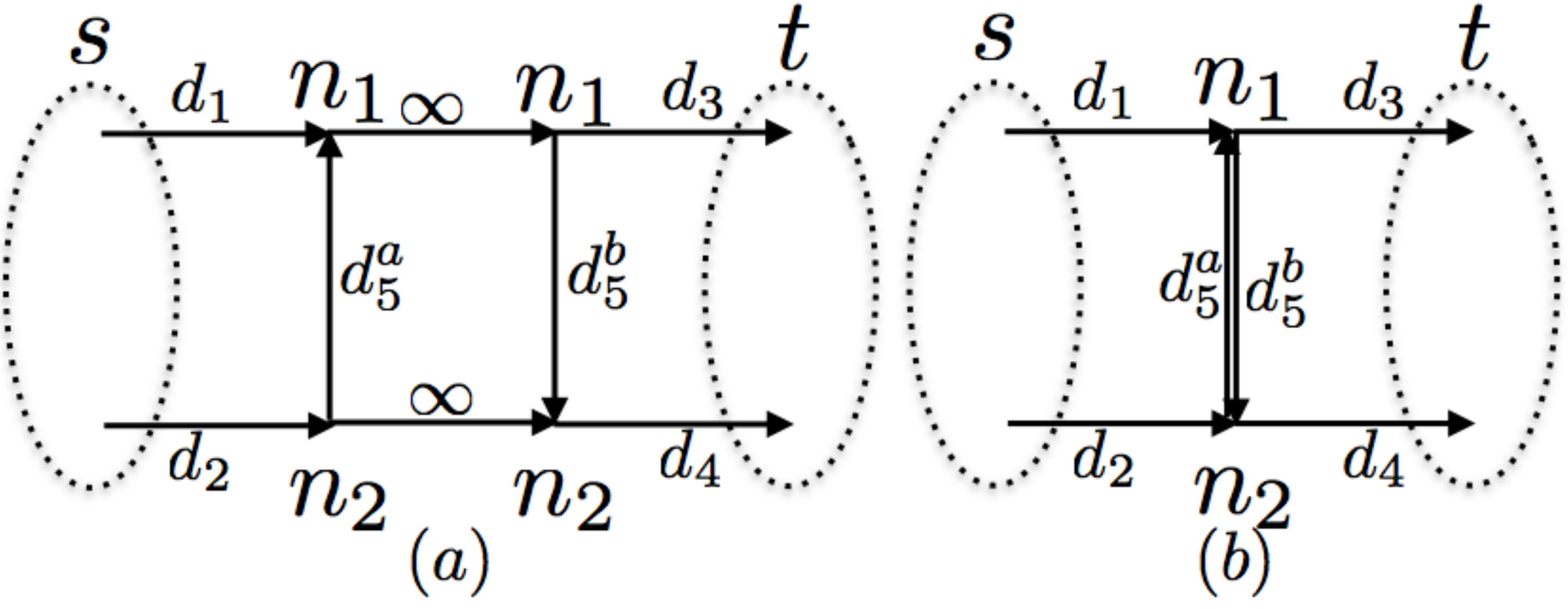}
  \caption{$(a)$ One of the equivalent networks of Fig.~\ref{fig3}, where the the arrow direction of the edge with dimension $d_5^a$ is chosen to be up, and the the arrow direction of the edge with dimension $d_5^b$ is chosen to be down. $(b)$ An equivalent network of (a), which corresponds to a graph with cycle (between the notes $n_1$ and $n_2$.}
  \label{fig4}
\end{figure}

In terms of quantum networks, one of the equivalent networks is shown in Fig.~\ref{fig4}. Again, since we allow unlimited classical communications between nodes, the choices of the direction of the edge with dimension $d_5^a$ and the edge with dimension $d_5^b$ are in fact arbitrary. For example, the network shown in Fig.~\ref{fig4}(a) is equivalent to the network shown in Fig.~\ref{fig4}(b), which corresponds to a graph with a cycle (between the nodes $n_1$ and $n_2$).

In case of quantum networks, different choices of the edge directions are all equivalent. However, this does have an effect in the corresponding classical network for the capacities $\CC(\tilde{\Net})$ and $\CCOne(\tilde{\Net})$, as we will discuss below. Since the graph of the network in  Fig.~\ref{fig4}(b) has a directed cycle of length $2$, Fig.~\ref{fig4}(a) demonstrates a way to transform it into an equivalent network with an acyclic graph. This then provides a general method to transform networks with directed graphs (with cycles of length $2$) into networks with directed acyclic graphs.

As an example, we consider the quantum repeater network given by Fig.~\ref{fig1}(a).
The corresponding tensor network is given by Fig.~\ref{fig2}. For
this network $\Net=(G,d,S,T)$, where $d$ is given by the five dimensions
$d_i$ ($i=1,2,3,4,5$), we fix $d_1=d_4=2$ and $d_2=d_3=3$. So the only
variable left is $d_5$, and we denote the corresponding network
by $\Net_{d_5}$.

Notice that in fact $\QMC(\Net_{d_5})=6$ for all $d_5>1$. And it is obvious that $\CROne(\Net_{6})=6$, which achieves the min-cut upper bound.

Based on the discussion of the relations between acyclic/cyclic networks, we can show that in fact we only need $d_5=4$ to achieve the upper bound, as shown in the following theorem.

\begin{theorem}
\begin{equation}
\CROne(\Net_{4})=6.
\end{equation}
\end{theorem}

\begin{IEEEproof}
We use the equivalent network $\tilde{N_4}$ in Fig.~\ref{fig4}(a), where $d_5^a = d_5^b = 2$. Since the corresponding directed graph $\tilde{\Net}_{4}$ is acyclic,
we have $\CROne(\Net_{4})=\CCOne(\tilde{\Net}_{4})$. So we only need to
find a classical network coding protocol that realizes
$\CCOne(\tilde{\Net}_{4})=6$.

We have the classical
protocol as follows. We denote the alphabet to be input into the channel labeled
by $d_1$ as $\{{0},{1}\}$, and the alphabet to be input into the channel labeled
by $d_2$ as $\{{0},{1},{2}\}$. Recall that $d_1=d_4=2$, $d_2=d_3=3$, and
$d_5=4$.

Now at the node $n_2$ with arrow up, we transmit one bit information from $n_2$
to $n_1$ depending on whether the input to the edge $d_2$ is $0/1$ or $2$. If it
is $0/1$, the node $n_2$ sends $0$ to $n_1$ through $d_5^a$, and both nodes $n_1$ and $n_2$ send their input to $t$. If it is $2$, then
the node $n_2$ sends $1$ to $n_1$ through $d_5^a$, node $n_1$ sends $2$ to $t$, and at the node $n_1$ with arrow down, one bit of information
is transmitted from node $n_1$ to node $n_2$ depending on the input to the edge $d_1$.
If the input is $0$, node $n_2$ sends $0$ to $t$; if the input is $1$, node $n_2$ sends $1$ to $t$.
\end{IEEEproof}

We remark that the classical network coding protocol that achieves the min-cut
upper bound as given in the proof above depends heavily on the fact that the network
of Fig.~\ref{fig4}(a) has a cycle. It is straightforward to check that any acyclic graph with $d_5=4$,
cannot achieve $\CCOne(\tilde{\Net}_{4})=6$. In fact, it is even not possibly achievable by any
acyclic graph with $d_5>4$, although the quantum network $\CROne(\Net_{4})=6$ holds
since the directions of the arrows do not matter in the quantum case. This reveals a subtle
difference between $\CCOne$ and $\CROne$.

\subsection{The gap between $\CROne(\Net)$ and $\CTOne(\Net)$}

To study the gap between $\CROne(\Net)$ and $\CTOne(\Net)$, we consider
the networks $\Net_{2}$ and $\Net_{3}$.

\begin{lemma}
\begin{equation}
\CTOne(\Net_{d_5})=6
\end{equation}
for all $d_5>1$.
\end{lemma}

\begin{IEEEproof}
For $d_5 = 2$, this can be verified by direct calculations. Then the statement for the general case $d_5 \geq 2$ follows as a consequence.
\end{IEEEproof}

However, there are indeed gaps between $\CROne$ and $\CTOne$, for some of
the networks $\Net_{d_5}$, as given by the following theorem.

\begin{theorem}
\begin{equation}
\CROne(\Net_{2})=\CROne(\Net_{3})=5.
\end{equation}
\end{theorem}

\begin{IEEEproof}
We first show that there exists a directed graph $\tilde{G}$, such that $\CCOne(\tilde{\Net}_{2})=5$. We choose the directed graph to be the one in Fig.~\ref{fig4}(b) with $d_5^a=d_5=2$, and $d_5^b=1$. That is, the direction of the edge $d_5$ is up. Now we denote the alphabet to be input into the channel labeled
by $d_1$ as $\{{0},{1}\}$, and the alphabet to be input into the channel labeled
by $d_2$ as $\{{0},{1},{2}\}$.

We then show that the input pairs of
$\{(0,0),(0,1),(1,0),(1,1),(0,2)\}$ can be transmitted from the source to the sink, via the
following protocol. At the node $n_2$, we transmit one bit information from $n_2$ to $n_1$ depending on whether the input to the edge $d_2$ is $0/1$ or $2$. If it
is $0/1$, the node $n_2$ sends $0$ to $n_1$ through $d_5$, both nodes $n_1$ and $n_2$ send whatever input to $t$. If it is $2$, then
the node $n_2$ sends $1$ to $n_1$ through $d_5$, $0$ to $t$, and node $n_1$ sends $2$ to $t$.

This shows that $\CROne(\Net_{3}) \geq \CROne(\Net_2) \geq 5$. That $\CROne(\Net_3)$ cannot go above $5$ can be verified by computer searches.
\end{IEEEproof}


\section{The network $(G,kd,S,T)$}

The following conjecture is given in ~\cite{cui2015quantum}.
\begin{conjecture}
\label{conj}
For any tensor network $\Net = (G,d,S,T)$, denote by $\Net^{\Mul k}$ the network $(G, kd,S,T)$. Then
\begin{equation}
\CTOne(\Net^{\Mul k})=\QMC(\Net^{\Mul k}),
\end{equation}
for sufficiently large $k>0$.
\end{conjecture}

\begin{figure}[htbp]
\centering
  \includegraphics[width=3.0in]{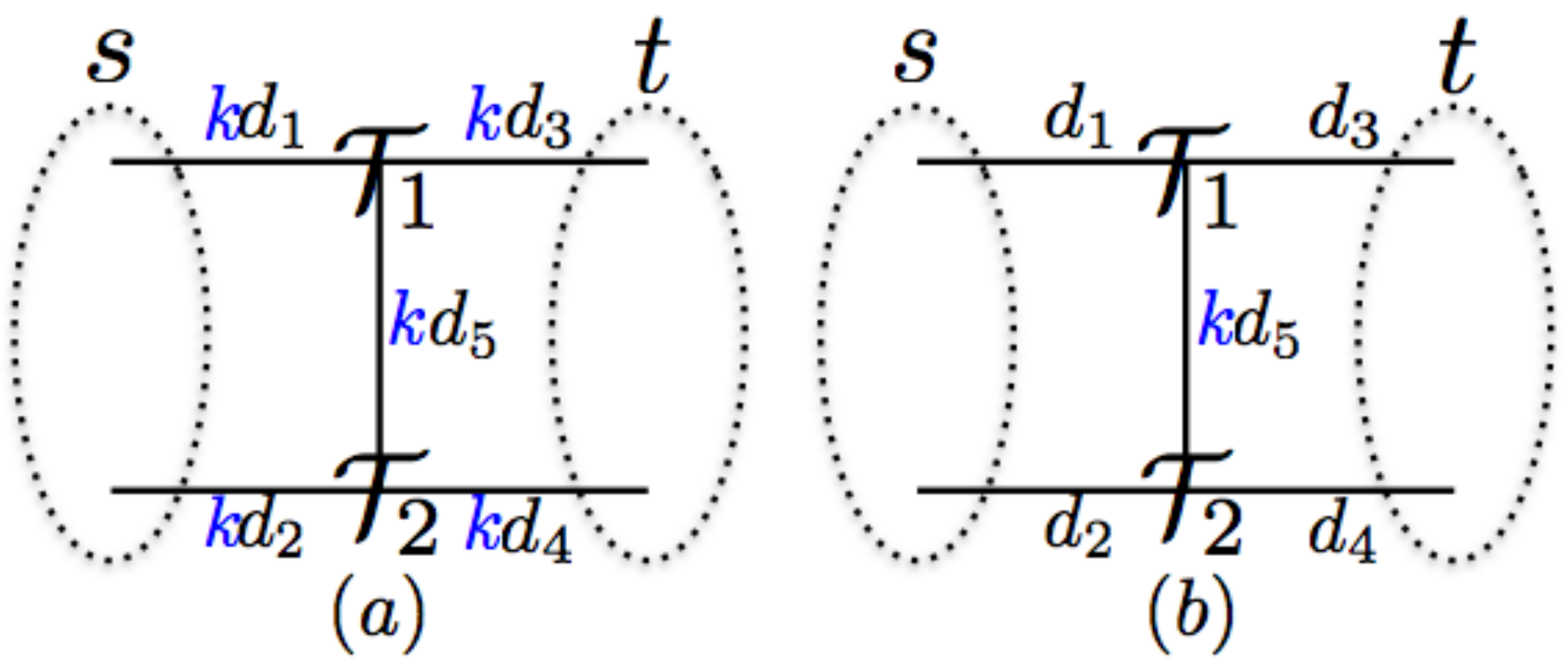}
  \caption{(a) Multiplying every edge of Fig.~\ref{fig1}(a)
  by a factor $k$. (b) The equivalent tensor network of Fig.~\ref{fig6}(b).}
  \label{fig5}
\end{figure}

In this section, we show that Conjecture~\ref{conj} holds for
the network given by Fig.~\ref{fig2}, for arbitrary choices
of $d_1,d_2,d_3,d_4$ and $d_5>1$. First,
notice that the tensor network corresponding to $\Net^{\Mul k}$
is given in Fig.~\ref{fig5}(a). We then show that it can be
reduced to the tensor network as given in Fig.~\ref{fig5}(b).

\begin{figure}[htbp]
\centering
  \includegraphics[width=3.0in]{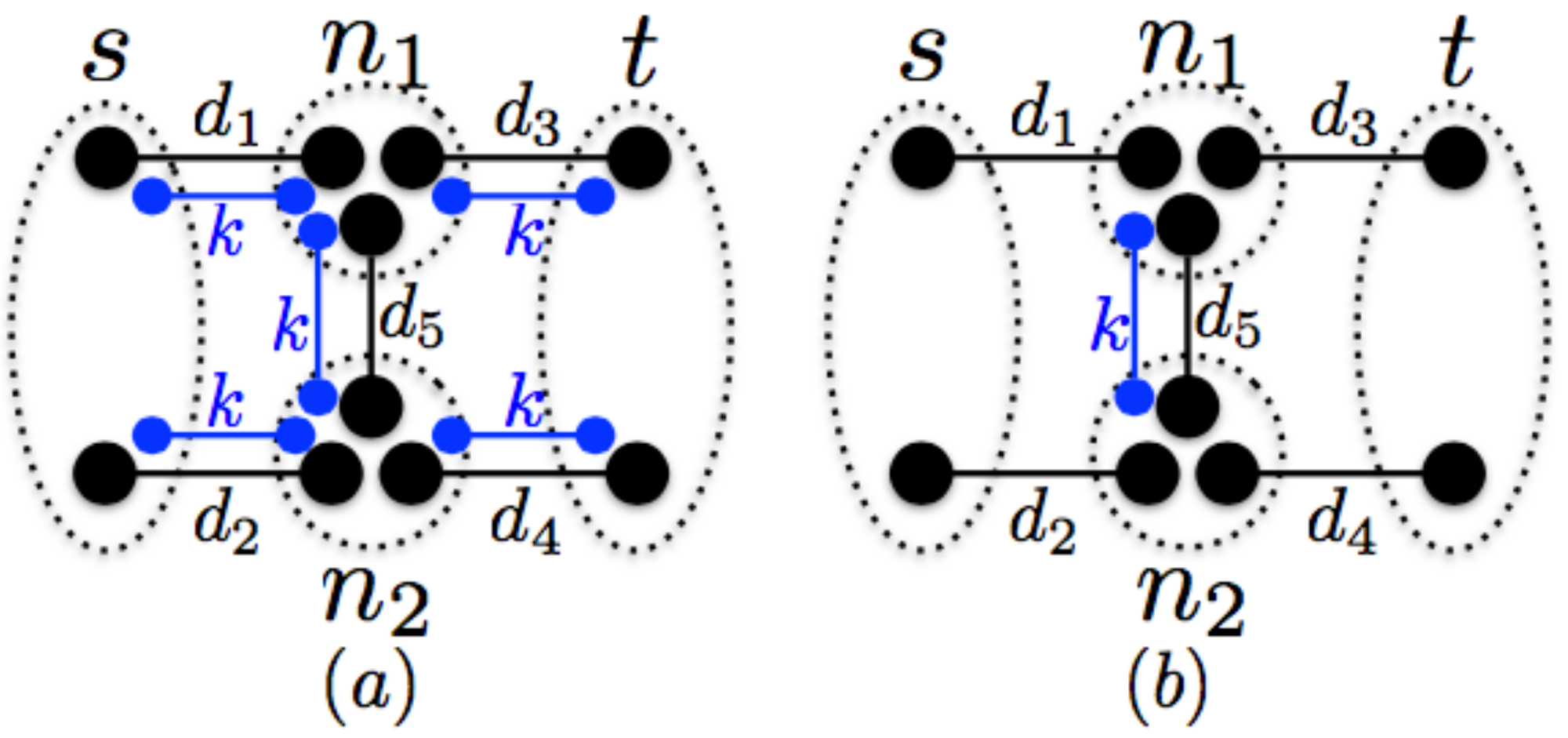}
  \caption{(a) The equivalent quantum repeater network of Fig.~\ref{fig5}(a). (b) The equivalent quantum repeater network of Fig.~\ref{fig5}(b).}
  \label{fig6}
\end{figure}

To see how this works, consider the equivalent quantum repeater
network of Fig.~\ref{fig5}(a), which is shown in Fig.~\ref{fig6}(a). Now
consider a quantum teleportation protocol for two qudits, each of dimension
$k$, from the source $s$ to the sink $t$. Now we teleport the first qudit
from $s$ to $t$ by the upper two maximally entangled states. Similarly,
we teleport the second qudit
from $s$ to $t$ by the lower two maximally entangled states. We are then left with
a quantum repeater network as given in Fig.~\ref{fig6}(b).

Since quantum teleportation is an LOCC protocol, it can surely be realized by an SLOCC protocol. Therefore, for the tensor network of Fig.~\ref{fig5}(a), we can transmit rank $k^2$ from $s$ to $t$,
with the residual network shown in Fig.~\ref{fig5}(b). For this tensor network with any given
$d_1,d_2,d_3,d_4$, sufficiently large $k$ will hence make the min-cut upper bound achievable by
$\CTOne$, based on a similar method as discussed in Sec.~\ref{sec:cycle}. Tensoring the two pieces together, we see that the original network achieves the min-cut upper bound.

Although the idea discussed above does not seem to suffice for proving Conjecture~\ref{conj} in the most general cases,
it may still be useful to study this conjecture for some other networks. We leave this as future work.

\section{Acknowledgement}  ZJ and
NY's research is supported by NSERC, NSERC DAS,
CRC, and CIFAR. BZ is supported by NSERC. Most of the present work was done through discussions among the authors via the interactive platform SciChat (www.scichat.com).



%

\bibliographystyle{IEEEtran}
\bibliography{Capacity}

\end{document}